# Effect of vanadium catalysts on hydrogen evolution from NaBH$_4$

Piotr Orłowski[1,2]* and Wojciech Grochala[1]*

1 Centre of New Technologies, University of Warsaw, Żwirki i Wigury 93, 02-089 Warsaw Poland

2 Faculty of Chemistry, University of Warsaw, Pasteur 1, 02-093 Warsaw Poland

* p.orlowski@cent.uw.edu.pl; w.grochala@cent.uw.edu.pl

Abstract:

NaBH$_4$ is a very cheap and hydrogen rich material and a potential hydrogen store. However, high temperature of its thermal decomposition (above 530°C) renders it inapplicable in practical use. Here, we have studied the effect of addition of diverse V-containing catalysts on thermal hydrogen desorption. It turns out that mechanochemical doping of NaBH$_4$ with vanadium metal, its oxides or nanoparticles lowers the temperature of pyrolysis significantly. Notably, NaBH$_4$ milled for 3 hrs with 25 wt.% of V$_2$O$_5$ or VO$_2$ releases ca. 70 % of stored hydrogen in the temperature range of *ca.* 370°C–450°C. On the other hand, precursors and solvents used to prepare rather uniform vanadium nanoparticles (~4nm) suspended in THF or less uniform and larger ones (~28nm) in o-difluorobenzene, have adverse effect on purity of hydrogen evolved.

1. Introduction.

Fossil fuels are widely used as energy source all over the world for centuries. Their intense usage led to environmental pollution and climate change, and that has forced humanity to seek alternative, green energy sources and carriers. Hydrogen, next to none in its high gravimetric energy density, is a clean source of energy, its combustion resulting only in water. This makes it a promising alternative to both fossil energy carriers as well as electric energy accumulators (1) (2) (3) (4) (5).

Sodium borohydride is one of the cheapest and richest in hydrogen compounds known. Therefore, it is not surprising that much effort was devoted to possibility of using it as a hydrogen storage material. Hydrogen can be released from NaBH$_4$ either during its hydrolysis or upon a thermal decomposition. Hydrolysis reactions in alkaline aqueous solutions take place at room temperatures upon contact with a catalyst (so called Hydrogen-on-demand® technology (6)). However, reversibility of this process is poor, as it is associated with necessity to recycle very stable oxoborates(III) back to BH$_4^-$ species. On the other hand, pyrolysis profile has a maximum desorption peak above 500°C (7) (values up to 534 ± 10°C at 1 bar of hydrogen (8) are given in the literature), which is quite high for mobile applications. One widely investigated solution to that problem is catalytic enhancement of pyrolysis. Variety of dopants has been studied by numerous teams, and we have grouped them for convenience into four main categories (Tables 1–4). I.e.: (I) first-transition metal (TM) period elements and their compounds, (II) second-TM period elements and their compounds, and (III) lanthanides and their compounds, and (IV) other hydrogen storage materials. Aside from that, composites of NaBH4 with diverse "inert" materials such as graphene, zeolites, etc. have been prepared and their thermal profiles studied (Table 5). Last but not the least, attempts to prepare NaBH$_4$ nanoparticles were done to decrease thermal decomposition temperature of the pristine material (Table 6).

Some of the most spectacular results achieved via doping with the first TM period species (Table 1) were seen for zinc(II) fluoride and titanium(III) chloride which evolve hydrogen around 95 and 100°C respectively. However, as typical for rather low temperature of thermal decomposition, $H_2$ gas is in these cases severely polluted with $B_2H_6$ or $B_2H_6$ and diglyme. For metals from the second TM period (Table 2) the lowest temperature of thermal decomposition of 379°C was achieved for 10 and 15 %$_{mol}$ doping with $NbF_5$. The third TM row catalysts were also tested but the decrease of thermal decomposition temperature was not impressive ($TaCl_5$ - 460°C and $ReCl_3$ - 465°C (9)). In lanthanide series (Table 3) the best result was achieved for neodymium fluoride doping (413°C).

*Table 1. Influence of the first transition metal-row based dopants on thermal decomposition temperature of $NaBH_4$ (5%$_{mol}$ doping if not specified differently)*

| Sc (10) | Ti | V | Cr (10) | Mn (10) |
|---|---|---|---|---|
| $ScF_3$ 501°C | Nano - 493°C (9) <br> $TiF_3$ - 300°C (11) <br> 5%$_{mol}$ $TiF_3$ - ~300°C (12) <br> $TiF_4$ - 522°C (10) <br> $TiCl_3$ 1:6 - 100°C * (13) <br> (*polluted with $B_2H_6$ and diglyme) <br> $TiB_2$ - 503°C (9) <br> $TiSiO_4$ - 489°C (9) | $VF_4$ - 499°C (10) <br> $NaBH_4$+$PrF_3$+$VF_3$ 3:1:0.2 - 417°C (14) <br> $V_3B$ - 478°C (9) | $CrF_3$ - 500°C | $MnF_3$ - 483°C |
| Fe (10) | Co | Ni | Cu (10) | Zn |
| $FeF_3$ 498 | $CoF_3$ - 501°C (10) <br> $Co_3B$ - 480°C (9) | Nano - 483°C (9) <br> $NiF_2$ - 452°C (10) <br> $NiF_2$ - 453°C (9) <br> wt. 10% Ni on Si/$Al_2O_3$ - 449°C (9) <br> $NiCl_2$ - 460°C (9) <br> $Ni_3B$ - 462°C (9) | $CuF_2$ - 476°C | $ZnF_2$ - 504°C (10) <br> $ZnF_2$ 1:2 - 95°C** (15) <br> (**polluted with $B_2H_6$) |

*Table 2. Influence of the second transition metal row-based dopants on thermal decomposition temperature of $NaBH_4$ (5%$_{mol}$ doping if not specified differently)*

| Y | Zr (10) | Nb (10) | Rh (9) | Pd (9) | Ag (10) | Cd (10) |
|---|---|---|---|---|---|---|
| $YF_3$ - 513°C (10) <br> $YF_3$ 3:1 - 423°C (16) <br> (x)$YF_3$+(1-x)$GdF_3$ 3:1 (17): <br> x=2/3 - 449°C <br> x=1/2 - 446°C <br> x=1/3 - 440°C | $ZrF_4$ - 503°C | $NbF_5$: <br> 2%$_{mol}$ - 442°C <br> 10%$_{mol}$ - 379°C <br> 15%$_{mol}$ - 379°C | Rh (5 wt %) on $Al_2O_3$ - 476°C | Nano - 420°C | $AgF$ - 498°C | $CdF_2$ - 512°C |

*Table 3. Influence of the lanthanide series metal based dopants on thermal decomposition temperature of $NaBH_4$ (5%$_{mol}$ doping if not specified differently)*

| La (18) | Ce (10) | Pr (14) | Nd (19) | Ho (20) |
|---|---|---|---|---|
| $LaF_3$ 3:1 - 396°C <br> La 3:1 - 442°C | $CeF_3$ - 506°C <br> $CeF_4$ - 502°C | $PrF_3$ 3:1 - 439°C | $NdF_3$ 3:1 - 413°C | $HoF_3$ 3:1 - 443°C |

*Table 4. Influence of co-milling of other hydrogen rich compounds with $NaBH_4$ on its thermal decomposition temperature (molar ratio was used if not described differently)*

| $MgH_2$ | $CaH_2$ (21) | $NaNH_2$ | $LiAlH_4$ (12) | $Li_3AlH_6$ (22) | $Ca(AlH_4)_2$ (23) |
|---|---|---|---|---|---|
| 1:2 - 420°C (24) <br> 1:2 with 5%$_{mol}$ of (25): <br> $TiF_3$ - ~470°C <br> $TiO_2$ - ~470°C <br> Zr - ~470°C <br> Si - ~470°C <br> BCC - ~470°C | 1:6 - 390°C | 1:1 - 330°C (26) <br> (polluted with ammonia) <br> 2:1 with Co-Ni-B catalyst (27): <br> 1 wt.% - ~300°C <br> 3 wt.% - ~285°C <br> 5 wt.% - 285°C <br> 7 wt.% - ~285°C | 1:1: <br> No catalyst - 446°C <br> 5%$_{mol}$ $TiF_3$ - ~300°C | 1:1 (1h) - 400°C <br> 1:1 (24h) - 392°C <br> 1:2 - 430°C <br> 1:3 - 430°C | 2$NaAlH_4$ + $Ca(BH_4)_2$ with 5 wt.% of $TiF_3$ - 400°C |

*Table 5. Thermal decomposition of NaBH$_4$ in its composite materials (molar ratio was used if not described differently)*

| graphene | carbon | fluorographite (28) | zeolites (29) | Ah-BN (30) | NaBF$_4$ (23) |
|---|---|---|---|---|---|
| wt.% 10 - 484°C (31) <br> ? - 426°C (32) | 20 wt.% of scaffolds CMK-3 - 235°C and 380°C (33) <br> (polluted with ammonia) <br> ~25wt.% of HSAG-500 (34): <br> infiltrated - ~270°C <br> melted - ~300°C <br> mixed - ~320°C | wt.% 45 - 141°C | MCM-22: <br> 1:1 mass - 490°C <br> 1:2 mass - 503°C <br> 1:3 mass - 507°C <br> 1:4 mass - 508°C <br> SAPO-34: <br> 1:1 mass - 483°C <br> 1:2 mass - 491°C <br> 1:3 mass - 493°C <br> 1:4 mass - 495°C | 1:1 mass - 399°C | 1:10 - 468°C <br> 1:2 - 305°C <br> (polluted with BF$_3$ and B$_2$H$_6$) |

*Table 6. Influence of nanosizing of NaBH$_4$ on its thermal decomposition temperature*

| Solvented ion stabilization (35) | Evaporation with 10%$_{mol}$ of ligands or solvents (36) | Anti-solvent precipitation | Graphene coated (32) |
|---|---|---|---|
| LiCl - 495°C <br> MgCl$_2$ - 445°C <br> NaI - 485°C <br> (all polluted with buthylamine) | Hexyloamine - 489°C <br> dodecyloamine - 476°C <br> octodecyloamine - 473°C <br> tertabuthylamonium bromide - 464 and 482°C <br> tetraoctylamonium bromide - 484°C <br> tetradecyloamonium bromide - ~500°C <br> dodecane - 478°C <br> dodecanothiol - 464°C <br> tri-decylic acid - ~400 and 474°C <br> tetrabutylphosphonium bromide - 471°C | Bare - 460 and 535°C (37) <br> Coated with: <br> Ni - 418°C (37) <br> Co - 350°C (38) <br> Cu - 400°C (38) <br> Fe - 380°C (38) <br> Sn - 450°C (38) | Ultrasonicated and dried - 400°C |

Co-milling NaBH$_4$ with other hydrogen storage materials (Table 4), particularly protic ones, often leads to facile evolution of H$_2$ from a composite system. As for NaBH$_4$, particularly promising results were achieved for 5 wt. % Co-Ni-B doped 2NaNH$_2$/NaBH$_4$ mixture, milled in hexane environment, for which temperature of pyrolysis with H$_2$ evolution drops to value as low as 285°C. However, hydridic-protic stores rarely show sufficient reversibility (39) (40).

If composites with diverse "inert" materials are considered (Table 5), substantial (45 wt.%) fluorographite addition resulted in the thermal decomposition point of 141°C; other composites usually yielded less appealing results.

The last group of proposed NaBH$_4$ modifications is either nanostructurization or coating it with metal, or graphene layers (Table 6) which can lead to decomposition temperature as low as 350°C for NaBH$_4$ nanoparticles coated with cobalt.

It may be seen from the literature screening that vanadium compounds have rather seldom been used to modify the H$_2$-release properties NaBH$_4$; and only moderate results were achieved by using VF$_3$, VF$_4$ or V$_3$B as catalysts. This is quite disappointing in view of the fact that from 40 to 100°C the γ phase of vanadium (II) hydride is known to exist in equilibrium with hydrogen gas and its metallic β form (41). This implies a rather low temperature of H$_2$ release from VH$_2$, especially in an H$_2$-free atmosphere. Moreover, the process of H$_2$ release from VH$_2$ must have a rather low activation barrier, as it tends to be fast. Consequently, vanadium metal or its compounds (reduced in situ to VNPs) could act as catalysts and destabilize [BH$_4$]$^-$ anions present in NaBH$_4$, thus decreasing the temperature of thermal decomposition of this material and leading to facile H$_2$ evolution at low temperature. Therefore, in this study we have researched doping of sodium borohydride with vanadium, vanadium oxides (V$_2$O$_3$, VO$_2$, V$_2$O$_5$) or vanadium nanoparticles and their effects on thermal stability of NaBH$_4$ hydrogen storage material.

2. Experimental
    2.1. Materials

NaBH$_4$ (≥98%), LiAlH$_4$ 2.0M solution in THF and catalysts: V$_2$O$_5$ (≥99,6%), VO$_2$ (≥99%) and V$_2$O$_3$ (99,99%) were purchased from Sigma Aldrich. Ph$_4$PCl (98%) used in Ph$_4$PBH$_4$ synthesis was delivered by abcr GmbH. Solvents: dichloromethane (DCM) (≥99,8% ≤50 ppm H$_2$O) and tetrahydrofuran (THF) (≥99,8%, ≤50 ppm H$_2$O) were purchased from ROTH, o-difluorobenzene (o-DFB) (98%) was obtained from Fluorochem. Before synthesis DCM and THF were dried with P$_2$O$_5$ and distilled; o-DFB was degassed using Schlenk line and LN$_2$. All solvents were then sealed in containers with molecular sieves.

2.2. Equipment

An Vigor SG1200/750TS-F glovebox was used for storing and preparing the samples; all reactions were performed in an inert gas (Ar) atmosphere (<10 ppm O$_2$, <0,05 ppm H$_2$O) or using Schlenk line. Mechanochemical doping was conducted in a stainless-steel disc bowl using a Testchem vibration mill (1400 RPM). Suspensions of VNPs were firstly sonicated using Polsonic Sonic-6D in an ice-cooled water bath. Thermal decomposition analysis of the samples was investigated using a thermogravimeter (TGA) combined with differential scanning calorimeter (DSC) from Netzsch STA 409 PG using Al$_2$O$_3$ crucibles. TGA/DSC chamber was constantly purged at a constant Ar (99.9999%) flow rate of 80 ml min$^{-1}$. The evolved gases were analyzed with a quadrupole mass spectrometer (MS) QMS 403 C (Pfeiffer Vacuum), connected to the TGA/DSC device by quartz capillary preheated to 200 °C to avoid condensation of low-boiling volatiles. Fourier Transform Infrared spectroscopy (FTIR) of all solid products of doping was measured using a Vertex 80v FT-IR spectrometer (Bruker). Anhydrous KBr was used as a pellet material.

2.3. Synthesis and doping

VCl(Al(OC(CF$_3$)$_3$)$_4$)$_2$·5ACN were synthesized according to (42). Vanadium nanoparticles (VNPs) were synthesized via reduction of vanadium (III) compounds. THF suspension was prepared using vanadium (III) chloride and THF solution of LiAlH$_4$. 10 ml of stoichiometric amount of reducer solution was added dropwise to 30 ml of 0.05 M solution of VCl$_3$ under intense stirring. In case of VNPs suspended in o-DFB, acetonitrile coordinated vanadium (III) bis(tetra(nonafluoro-tert-butoxy)aluminate) monochloride was used as source of vanadium and tetraphenylphosphonium borohydride acted as reducer. 20 ml of 0.015 M reducer solution was added dropwise to 20 ml of 0.005M solution of vanadium salt, under intense stirring. Excess of reducer was used to ensure complete reduction of vanadium cation. Both suspensions were left for 24h. Attempt of purification by centrifuging brought no results since VNPs could not be separated from their suspension that way and were used without further purification. 5 or 25 w.t.% of powder catalysts (V, V$_2$O$_3$, VO$_2$ or V$_2$O$_5$) were mixed with NaBH$_4$ and milled in 100g batches for 5 or 30 min in milling periods of 1 and 5 min respectively, altered with LN$_2$ cooling to avoid overheating of samples. As-prepared were then tested with TGA/DSC in 20-450 or 20-600°C range with 5 K/min heating rate together with MS of evolved gases and compared with milled pure NaBH$_4$ as reference. 100mg of NaBH$_4$ was doped with VNPs via 20 min sonication in ice-cold bath with 2ml or 10 ml of THF and O-DFB VNPs suspension respectively, then dried using Schlenk line and further milled in the same way as powder-doped samples.

3. Results
3.1. Synthesis of VNPs

TEM analysis of synthesized VNPs (Fig. 1) shows that nanoparticles prepared in THF using LiAlH$_4$ are smaller and more uniform than those synthesized in o-DFB. Roughly calculated mean size of VNPs are 28 ± 9.2 nm and 4 ± 1.7 nm for o-DFB and THF suspensions respectively. Since LiAlH$_4$ is a much stronger reducer than Ph$_4$PBH$_4$, vanadium is reduced faster and therefore nanoparticles grow to a smaller size. Unstirred VNPs are stable and stay suspended without any sign of agglomeration for months, despite no surfactant being used.

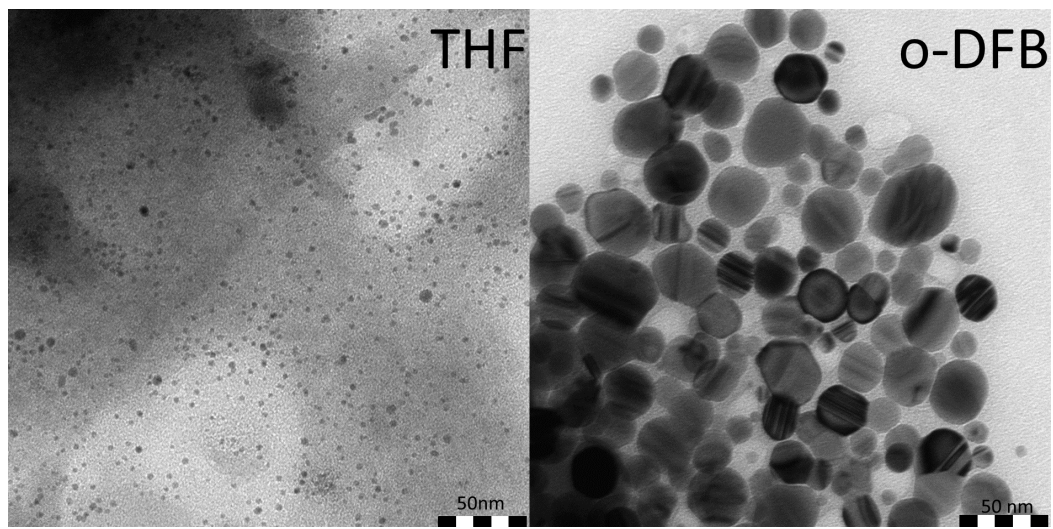

*Figure 1. TEM images of VNPs prepared in THF and o-DFB solutions*

3.2. Thermogravimetric and spectroscopic analysis of doped samples

To screen behavior of diverse V dopants, samples with 5 wt.% of catalyst milled in 5 short periods of 1 min each were preliminarily studied. We label them here as Na:X5%_5min where X stands for catalyst used. The TG-DSC profiles together for these samples as well as a reference sample of pristine $NaBH_4$ (milled in a similar way) together with ion current for diverse mass peaks in the MS of the evolved gas, are shown jointly in Fig.2. The picked masses in MS correspond to hydrogen, water and diborane.

There is a relatively minor effect of these short-milled dopants on the thermal decomposition of $NaBH_4$, with noticeable effect seen only for $VO_2$ (inflection of the TG curve at 402°C associated with mass loss of 0.6 wt. % up to 450°C) and a more pronounced one for $V_2O_5$ (clear mass decrease above 398°C, associated with the total mass loss of 1.0 wt. % up to 450°C). MS clearly indicates that these events correspond to endothermic $H_2$ evolution but not associated by water or diborane.

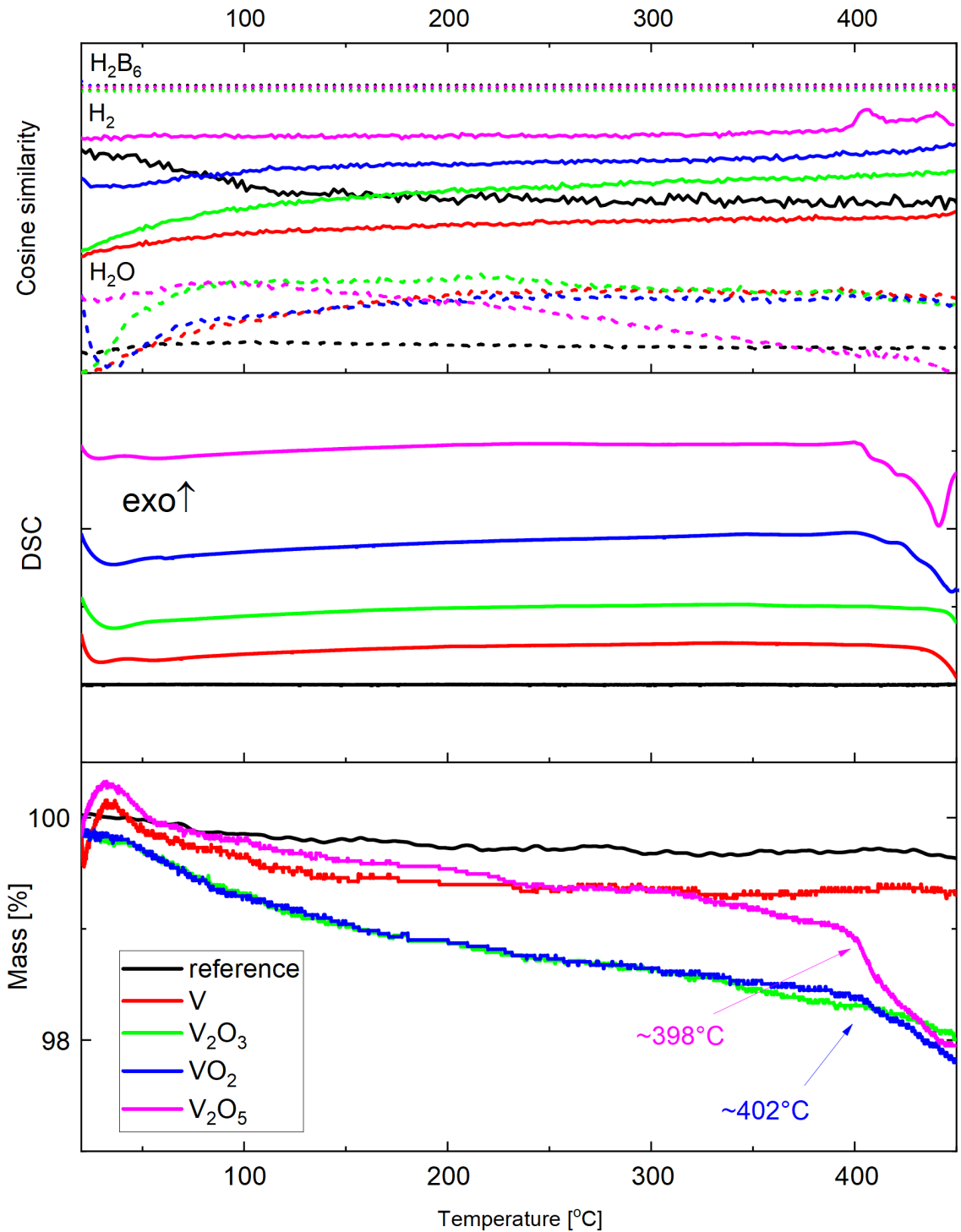

*Figure 2. TGA/DSC + MS curves of pristine and doped for 5 min with 5 wt.% of catalyst $NaBH_4$. DCS and MS curves are stacked in each section with offset applied for their better distinction, but without distortion of quantitative proportion between them. The fact that the TGA profiles are not flat for the doped samples may be associated with absorption of a very small amount of moisture during milling; water vapour is then released during heating. Note that the TGA scale comprises of 3 wt.% only, hence the noise is more pronounced than in all forthcoming figures.*

Encouraged by this preliminary result, we have further focused on $V_2O_5$-doped samples. Another three samples with increased content of vanadium pentoxide of 25 wt.% were prepared via 5, 30 and 180 min milling and referred to as Na:$V_2O_5$25%_5min, Na:$V_2O_5$25%_30min and Na:$V_2O_5$25%_180min, respectively. As obtained samples were inspected with FTIR and results were compared to those for pristine NaBH$_4$ and V$_2$O$_5$ (Fig. 3).

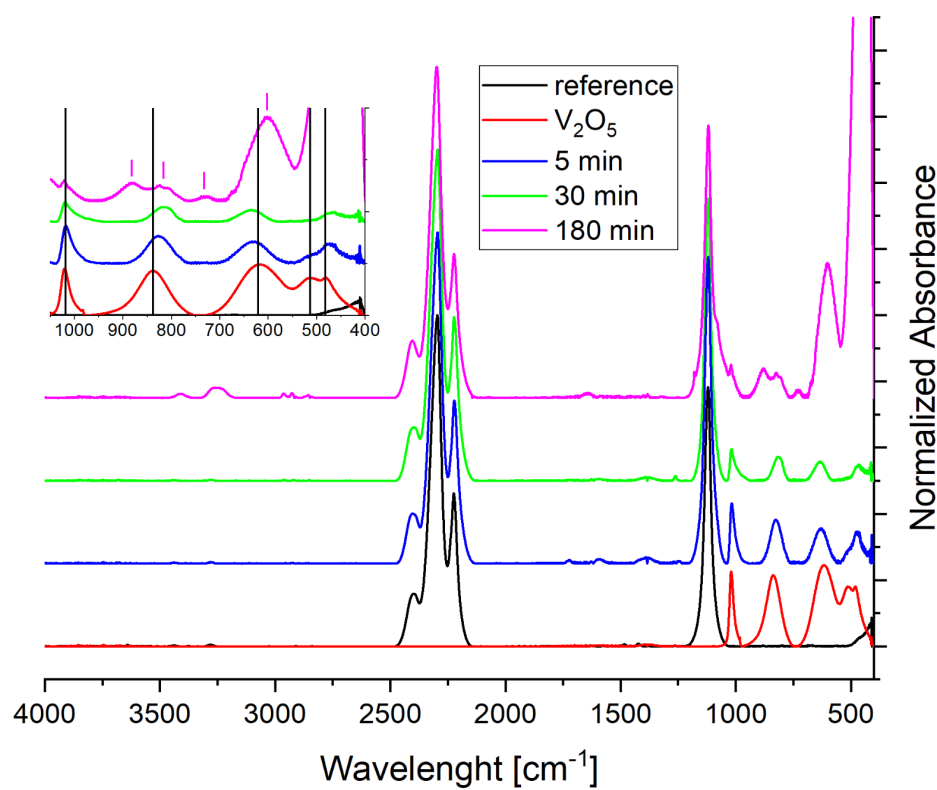

*Figure 3. FTIR spectra of Na:V$_2$O$_5$25%_5min and Na:V$_2$O$_5$25%_30min in comparison to pure NaBH$_4$ and V$_2$O$_5$. Spectra of NaBH$_4$-containing species were normalized to the most intense peak originating from B-H stretching. Spectra were shifted in vertical scale for clarity. Inset shows a blow of the 400–1000 cm$^{-1}$ range.*

No change of wavenumber at peak, nor of relative intensities of diverse absorption bands, characteristic for NaBH$_4$ (2400, 2295, 2223 and 1120 cm$^{-1}$) was observed upon doping with V$_2$O$_5$, independent of the milling time. However, bands at 1018, 838, 621 and 482 cm$^{-1}$ which are characteristic for V$_2$O$_5$ clearly decrease their intensity if the sample is milled 30 instead of 5 min and one at 514 cm$^{-1}$ can be barely seen already in a 5 min milled sample. Noticeable shifts in 838 and 621 cm$^{-1}$ peaks can be observed which progresses in function of milling time. If milled for 3 hours, those bands disappear almost completely and few new bands can be described. Very weak one at 732 cm$^{-1}$, two weak at 882 and 816 cm$^{-1}$, moderate at 602 cm$^{-1}$ and very strong band at 436 cm$^{-1}$ (off the scale) together with bands characteristic for water in the range of 3000-3600cm$^{-1}$. This result indicates that V$_2$O$_5$ is slowly reacting with milled borohydride and undergoing a progressive reduction, which is associated with the formation of the unknown species. In each case, however, there remains a lot of bulk NaBH$_4$ in these samples as evidenced by a stable BH-stretching and HBH bending region.

All three samples were studied with TGA-DSC-MS and results were compared to those for Na:V$_2$O$_5$5%_5min obtained at very short milling time (5min) and low catalyst load of 5 wt.% (Fig. 4).

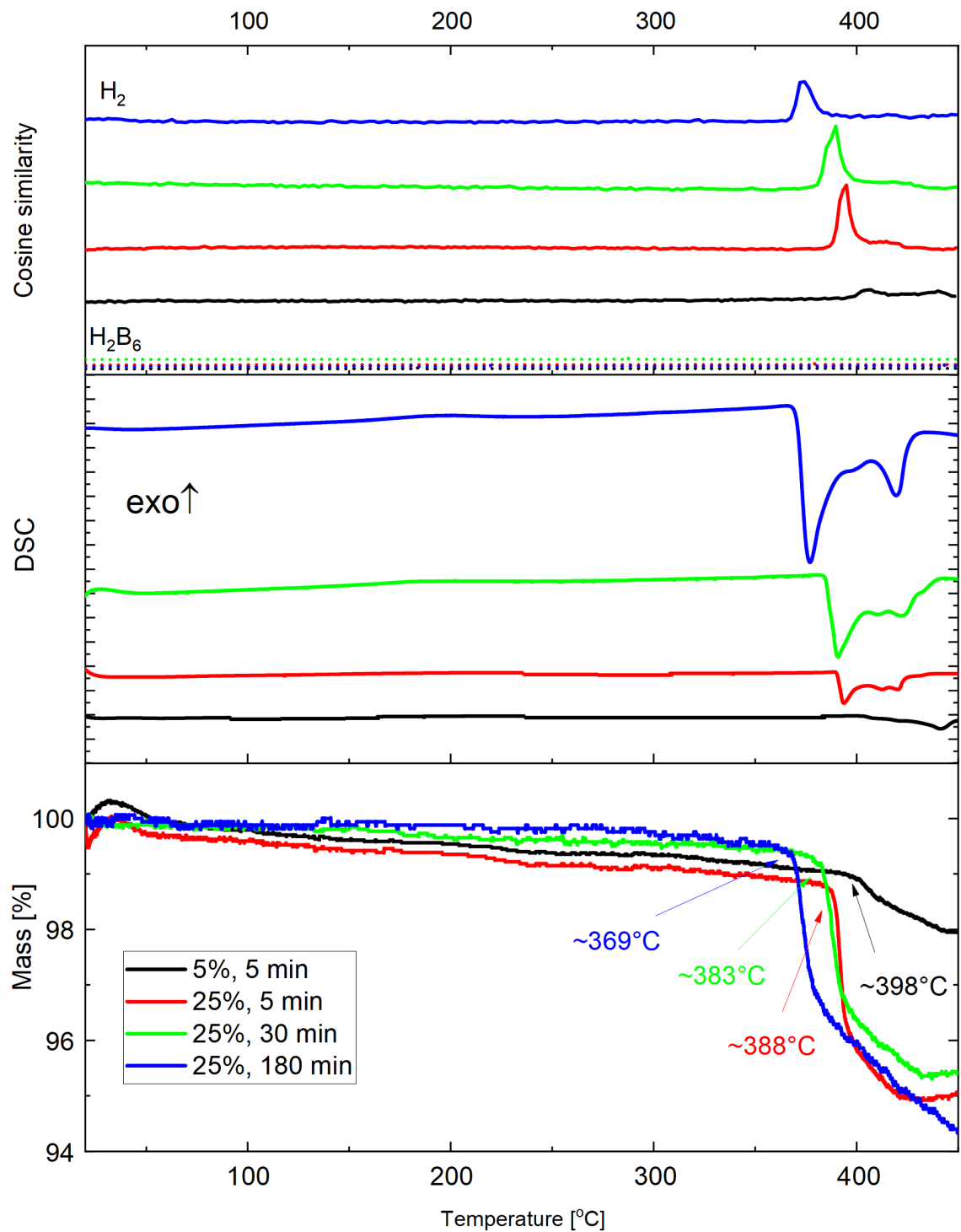

*Figure 4. TGA/DSC + MS curves of Na:V$_2$O$_5$5%_5min, Na:V$_2$O$_5$25%_5min, Na:V$_2$O$_5$25%_30min and Na:V$_2$O$_5$25%_180min samples. DCS and MS curves are stacked in each section with offset for visual improvement, without proportion distortion.*

There is a clear effect of both the catalyst load and a milling time on thermal decomposition of NaBH$_4$/V$_2$O$_5$ conglomerates as seen in TG-DSC-MS profiles. The weak effect seen for the Na:V$_2$O$_5$5%_5min sample at 398°C, is enhanced and shifted to lower temperature of 390°C for the sample Na:V$_2$O$_5$25%_5min and further shifted to 383°C for the sample Na:V$_2$O$_5$25%_30min. The lowest onset of thermal decomposition is seen for the 3hrs-milled sample (369°C). All these events correspond to sharp and complex endothermic peaks of hydrogen desorption and are adjoined by sharp H$_2$ release in the temperature range of 380 and 410°C as seen in the MS. Hydrogen is free from major contamination, as additionally evidenced by the MS.

The total mass loss between the onset of thermal decomposition and 450°C is 4.0, 4.1% and 5.4% for 25%-doped samples milled 5, 30 and 180 min, respectively (Figure 5). The latter value corresponds to ca. 70% of total hydrogen stored by the sample. Importantly, as indicated by MS, hydrogen is not contaminated by common impurities such as B$_2$H$_6$, other boron hydrides, water, etc.

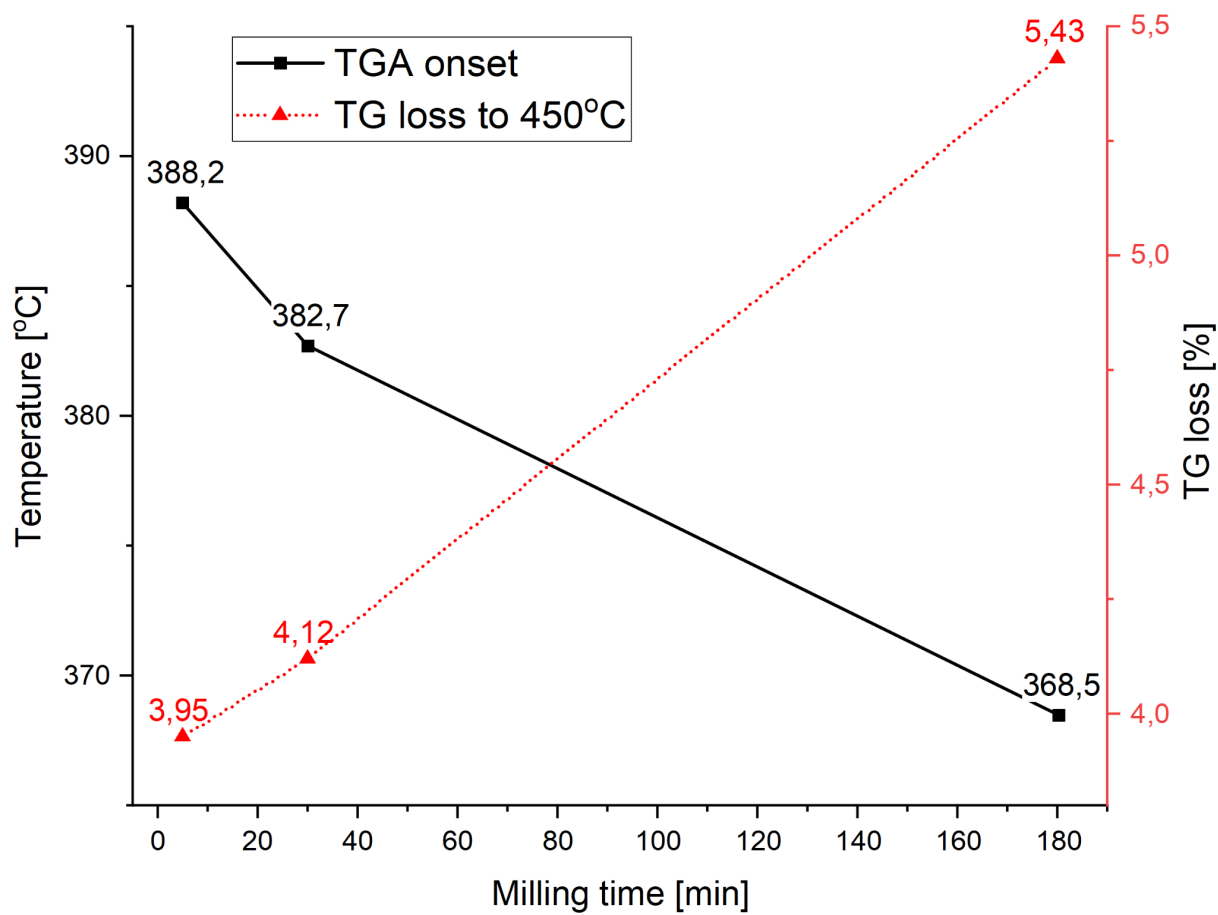

Figure 5. Mass loss up to 450°C observed for Na:V$_2$O$_5$25%_5min, Na:V$_2$O$_5$25%_30min, Na:V$_2$O$_5$25%_180min with respect to the milling time together with the onset of the thermal decomposition temperature.

Having achieved a marked reduction of the thermal decomposition temperature of NaBH$_4$ by high load of V$_2$O$_5$ and a prolonged milling time, we have applied the same procedure to the remaining vanadium oxides. Again, samples are labelled Na:X25%_30min, where X refers to the catalyst used. The TG-DSC-MS profiles for all samples were jointly shown in Fig. 6.

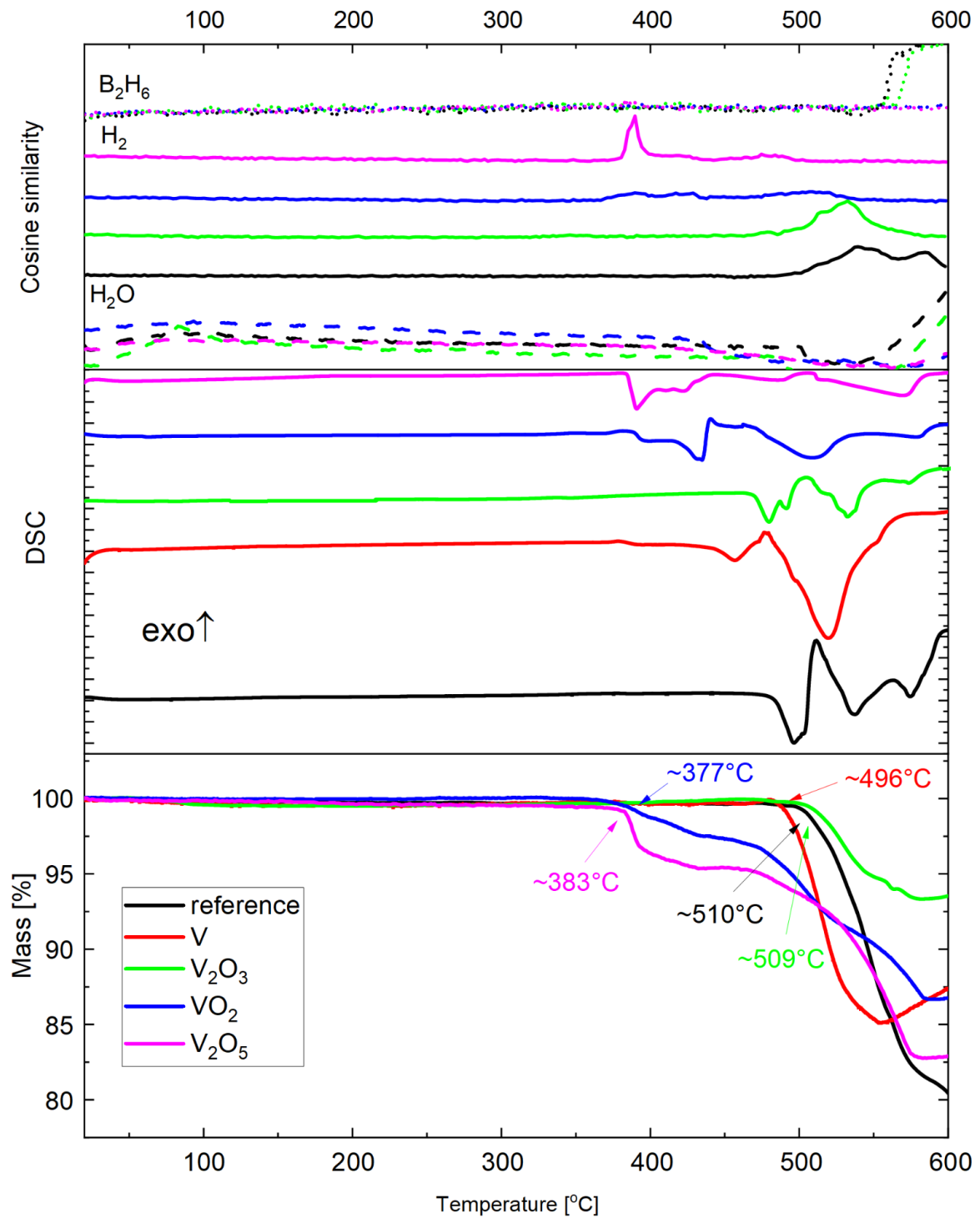

*Figure 6. TGA/DSC + MS curves of pristine NaBH$_4$ and sampled doped for 30 min with 25 wt.% of diverse vanadium catalysts. DCS and MS curves are stacked in each section with offset for better distinction, without distortion of the quantitative proportions.*

Substantial effect (and comparable to that for V$_2$O$_5$) on TG and DSC profiles is seen for and VO$_2$; here, the onset of thermal decomposition is around 377°C, but the amount of hydrogen evolved up to 450°C is smaller than for V$_2$O$_5$ additive. The effects seen for V$_2$O$_3$ and V are the least pronounced.

It is important to note that the total mass loss up to 600°C in almost all samples exceeds the total H contents of the samples. This is because sodium vapour is partly released at high temperatures (13) (37) (43). It is uncertain at this time what amount of sodium remains in the doped samples and which is chemical identity of the phase(s) containing this metal.

Having studied vanadium and vanadium oxides-doped samples, our attention turned to vanadium nanoparticles. Samples doped with vanadium nanoparticles were denoted Na:VNPs$_{THF}$ and Na:VNPs$_{O-DFB}$, and they were studied with TGA-DSC-MS analogously as previous samples (Fig. 7). Reference curves for milled NaBH$_4$ were added for comparison.

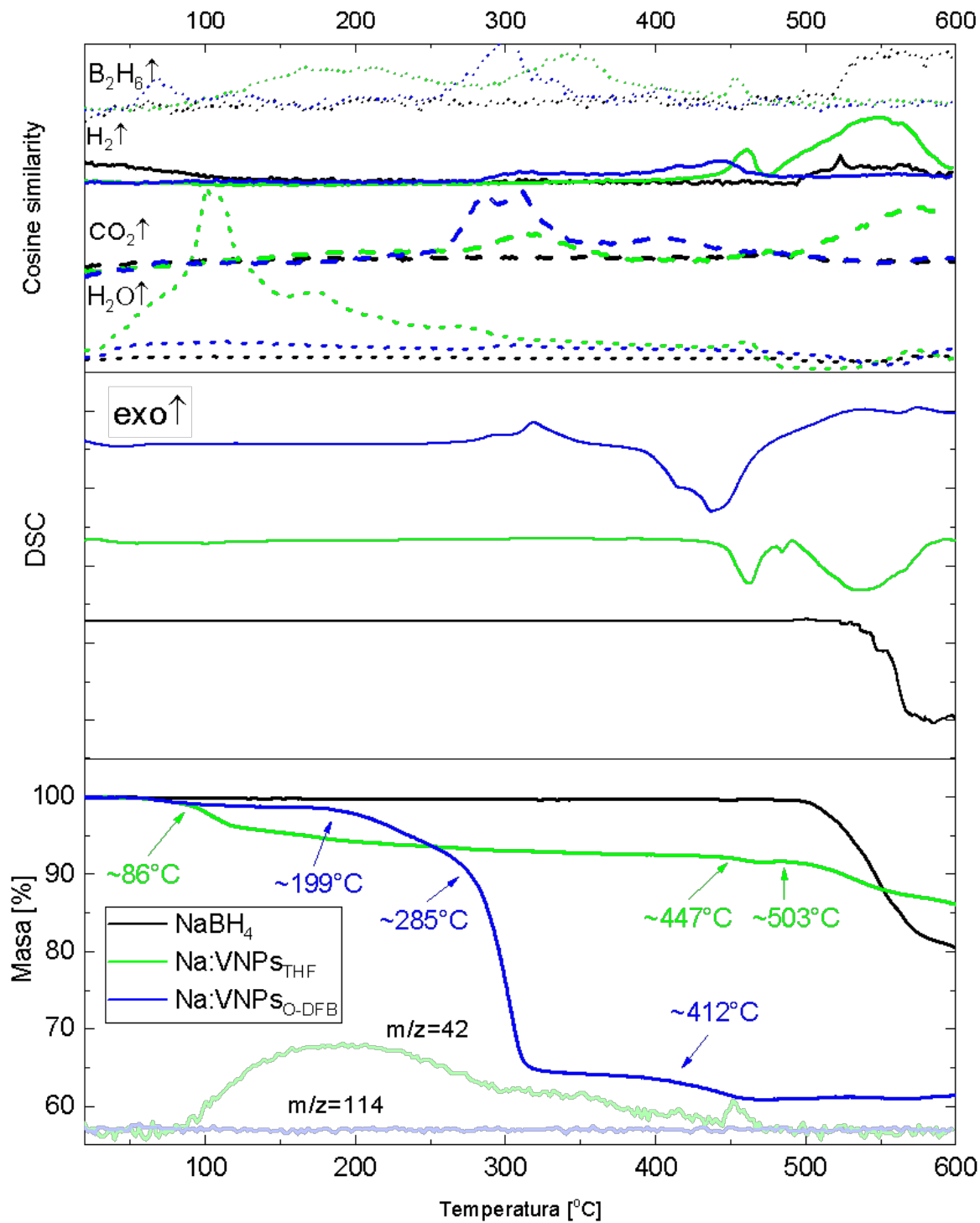

*Figure 7. TGA/DSC + MS curves of milled NaBH₄ and both VNPs doped samples. DCS and MS curves are stacked in each section with offset for better distinction, but without distortion of quantitative proportions. m/z =42 and 114 corresponds to THF and o-DFB MS main peaks respectively.*

Both VNPs-doped samples seem to release $H_2$ at a lower temperature than the reference sample, as may be judged from MS profiles. However, the doped samples also release copious amounts of $CO_2$, $B_2H_6$ and solvent residues (only in case of Na:VNPs$_{THF}$ ) at even lower temperatures, as seen in the MS and TGA profiles. These contaminations most probably come from byproducts of VNPs synthesis and are certainly highly undesired for any practical application.

4. Conclusions

We have studied the effect of vanadium and its oxides and nanoparticles (VNPs), on thermal decomposition of $NaBH_4$. It turns out that 3 hrs milling of $NaBH_4$ with 25 wt.% of $V_2O_5$ decreases temperature of $NaBH_4$ decomposition by as much as 165°C with respect to pristine sodium borohydride. $VO_2$ doping also decreases temperature of $NaBH_4$ pyrolysis, but to a lesser degree; moreover, hydrogen release peak is much broader than for $V_2O_5$ doping. The hydrogen released up to 450°C is free from $B_2H_6$, other boron hydrides, and water. In contrary to expectations, vanadium metal provides nearly no improvement in hydrogen desorption temperature of sodium borohydride. Doping of VNPs results in contamination of evolved $H_2$ with some precursors used for their synthesis, which is a highly undesirable effect.

Extension of these studies may focus on the chemical nature of vanadium in the long-milled and thermally decomposed samples, attempts to achieve at least partial reversibility of $H_2$ storage as well as broadening of the study towards other light metal borohydrides.

5. Acknowledgements

Authors acknowledge financial support from the Polish Ministry of Science and Higher Education of Poland, Diamentowy Grant, 501-D313-59-0557788.